\documentclass[twocolappendix]{aastex63}
\usepackage{amsmath}
\usepackage{amsfonts}
\usepackage{amssymb}
\usepackage{xcolor}
\usepackage{longtable}
\usepackage{makecell}
\usepackage{graphicx}
\graphicspath{{./Images/}}
\usepackage{hyperref}
\usepackage[T1]{fontenc}

\defcitealias{EHTC_2019_1}{EHTC M87~I}
\defcitealias{EHTC_2019_2}{EHTC M87~II}
\defcitealias{EHTC_2019_3}{EHTC M87~III}
\defcitealias{EHTC_2019_4}{EHTC M87~IV}
\defcitealias{EHTC_2019_5}{EHTC M87~V}
\defcitealias{EHTC_2019_6}{EHTC M87~VI}
\defcitealias{EHTC_2021_7}{EHTC M87~VII}
\defcitealias{EHTC_2021_8}{EHTC M87~VIII}
\defcitealias{EHTC_2023_9}{EHTC M87~IX}

\defcitealias{EHTC_2022_1}{EHTC SgrA~I}
\defcitealias{EHTC_2022_2}{EHTC SgrA~II}
\defcitealias{EHTC_2022_3}{EHTC SgrA~III}
\defcitealias{EHTC_2022_4}{EHTC SgrA~IV}
\defcitealias{EHTC_2022_5}{EHTC SgrA~V}
\defcitealias{EHTC_2022_6}{EHTC SgrA~VI}

\begin{document}

\title{Adiabatic Index in Fluid Models of Collisionless Black Hole Accretion}

\author[0000-0002-0393-7734]{Charles~F.~Gammie}
\affiliation{Department of Astronomy, Department of Physics, ICASU, and NCSA, University of Illinois at Urbana-Champaign}

\begin{abstract}

Models of highly sub-Eddington accretion onto black holes commonly use a single fluid model for the collisionless, near-horizon plasma.  These models must specify an equation of state.  It is common to use an ideal gas with $p = (\gamma - 1) u$ and $\gamma = 4/3, 13/9,$ or $5/3$, but these produce significantly different outcomes.  We discuss the origins of this discrepancy and the assumptions underlying the single fluid model.  The main result of this investigation is that under conditions relevant to low luminosity black hole accretion the best choice of single fluid adiabatic index is close to but slightly less than $5/3$.  Along the way we provide a simple equilibrium model for the relation between the ion-to-electron dissipation ratio and the ion to electron temperature ratio and explore the implications for electron temperature fluctuations in Event Horizon Telescope sources.

\end{abstract}


\section{Introduction}

Nearly all supermassive black holes in the centers of galaxies have luminosity far below the Eddington luminosity \citep{ho08}.  Examples include M31* \citep{li2009} and the Event Horizon Telescope (EHT) sources M87* \citep{EHTC_2019_1} and Sgr A* \citep{EHTC_2022_1}.  These low luminosity black holes likely accrete via an optically thin, geometrically thick disk \citep{yuan14}.  Phenomenological models imply that, near the horizon, the plasma is near the virial temperature and thus that the electrons are mildly relativistic.  In this regime the cross section for Coulomb scattering is of order the Thomson cross section.  Since low luminosity black hole accretion flows are Thomson-thin, the mean free path to Coulomb scattering is larger than the size of the system and the plasma is said to be {\em collisionless}. 

The fluid approximation can be formally justified only if the mean free path $\lambda$ is small compared to the system size $L$. \footnote{i.e. the Knudsen number ${\rm Kn} \equiv \lambda/L \ll 1$.}  In a magnetized plasma charged particles orbit helically around field lines, with gyroradius $r_g = (\beta_\perp \Gamma) (m c^2/(|q| B)) = 3.1 \times 10^6 (\beta_\perp \Gamma/B)$cm.  Here $B$ is field strength in Gauss, $q$ is particle charge, $\beta_\perp \equiv v_\perp/c$ is the component of particle velocity perpendicular to the field, and $\Gamma$ is particle Lorentz factor.  Then $r_g$ is the {\em effective} mean free path perpendicular to the field. Motion in the plane perpendicular to the field is thus fluid-like if $r_g \ll L$, a condition that is always satisfied for black hole accretion flows that are luminous enough to be observed. 

What is the mean free path parallel to the magnetic field?  In the absence of Coulomb scattering
the dominant scattering process is almost certainly wave-particle scattering, so the mean free
path is intimately linked to the wave spectrum.  There is now a growing understanding of kinetic
instabilities in disks and their ability to excite fluctuations that scatter particles
\citep[e.g.][]{Kunz:2014}.  Stochastic plasma echoes also suppress Landau damping and may be
able to ``fluidize'' collisionless turbulence \citep{meyrand2019}.  Nevertheless a complete
prescription for the effective mean free path in collisionless plasmas does not yet exist.  Future
progress is expected from global and local kinetic simulations \citep[e.g.][]{parfrey19,
bacchini22}.  In this paper we assume that the effective mean free path parallel to the magnetic
field is small enough that the plasma can be modeled as an ideal fluid.

Even if a collisionless plasma can be treated as a fluid it need not be fully relaxed: the ion and electron temperatures may differ.  To motivate this idea consider a marginally collisional plasma where  relaxation is dominated by Coulomb scattering.  The relaxation time is different for electron-electron, ion-ion, and ion-electron scattering.  It is shortest for  electron-electron scattering, longer by a factor of $(m_i/m_e)^{1/2}$ for ion-ion collisions, and longer by another factor of $(m_i/m_e)^{1/2}$ for ion-electron collisions \citep{spitzer56}.  Because the ion-electron relaxation time is relatively long the marginally collisional plasma can enter a two-temperature state in which the ions and electrons are separately relaxed by Coulomb scattering at different temperatures.  Two temperature plasmas were introduced in black hole accretion theory by \cite{shapiro76}. 

For the wave-particle scattering that likely controls relaxation in collisionless plasmas the characteristic wavelength and polarization of fluctuations that are most effective in scattering particles depends on particle charge to mass ratio.  One band and polarization of the fluctuation spectrum may be most effective at scattering ions and a second band and polarization may be most effective at scattering electrons. If there is no band/polarization that scatters both species and exchanges energy between them then a two-temperature state is a plausible outcome for a collisionless plasma as well. 

From here on we assume that ions and electrons are individually  relaxed with temperatures $T_i$ and $T_e$.  The temperature ratio
\begin{equation}
R \equiv \frac{T_i}{T_e}
\end{equation}
depends on each parcel of plasma's history of compressive heating, dissipation of kinetic and magnetic energy into the ions (heating rate $\equiv Q^+_i$),  electrons (heating rate $Q_e$), electron-ion energy exchange, and electron radiative cooling.  The dissipative branching ratio $Q^+_i/Q^+_e$ is not known but in most models $Q^+_i/Q^+_e > 1$, suggesting that  $R > 1$ \cite{quataert99, sharma07, howes10, kawazura19}.

At the densities found near low luminosity supermassive black holes, each component of the plasma behaves like an ideal gas.  The total pressure
\begin{equation}
    p = p_i + p_e = n_i k T_i + n_e k T_e
\end{equation}
and total internal energy
\begin{equation}\label{eq.sumu}
    u = u_i + u_e = \frac{1}{\gamma_i - 1} n_i k T_i + \frac{1}{\gamma_e - 1} n_e k T_e.
\end{equation}
Here $k$ is Boltzmann's constant, $\gamma_i, \gamma_e$ are the adiabatic indices of the ion and electron fluids, assumed constant, and $n_i, n_e$ are number densities, with $n_e = Z n_i$ (assuming a single ion species). The correct way to treat the thermodynamics of the resulting two-temperature plasma is to evolve separate energy equations for the ions and electrons \citep[e.g.][]{sadowski17, liska24}.   Notice that such a two-fluid model must introduce explicit models for heating, cooling, and energy exchange between the ions and electrons. 

In practice it is common to use a less computationally expensive single fluid model \citep[e.g.][]{EHTC_2019_5, porth19, EHTC_2021_8, EHTC_2022_5}, which does not require explicit models for electron and ion thermodynamics.  In a single fluid model the equation of state is $p = (\gamma - 1) u$, where $\gamma$ is a composite adiabatic index.  For a combination of physical and numerical reasons simulators have set $\gamma = 4/3, 13/9$, and $5/3$.  Profiles of the effective temperature $p/\rho$ differ significantly between models with $\gamma = 4/3$ and $\gamma = 5/3$ (V. Dhruv, B. Prather, private communication): models with $\gamma = 5/3$ have higher temperature close to the black hole than models with $\gamma = 4/3$.  So the choice of adiabatic index matters.

Which adiabatic index is correct?  The question is ill posed because a two-temperature plasma has (at least) one extra degree of freedom.\footnote{In reality ions and electrons will also have temperature anisotropies driven by compression and expansion of the plasma as in \cite{chandra_extended_2015, galish23}.}  It is possible, however, to ask which $\gamma$ best models certain features of the plasma. Section \ref{sec.therm} reviews thermodynamics of transrelativistic plasmas and computes the adiabatic index for various assumptions about ion-electron coupling.  Section \ref{sec.two} introduces a model for a collisionless plasma that explicitly includes compression, heating, cooling, and ion-electron energy exchange.  In Section \ref{sec.limits} we solve the model in the strong and weak electron-ion coupling limits and show that these recover the expressions found in Section \ref{sec.therm}.   If the temperature ratio is steady then the model enables us to explicitly compute the relationship between $Q_i/Q_e$ and $R$, which we do in Section \ref{sec.eqrat}.  Section \ref{sec.conc} sums up, concluding that if $R \gg 1$ then $\gamma$ slightly less than $5/3$ best models a collisionless plasma.  

\section{Thermodynamics and Equation of State Review}\label{sec.therm}

\subsection{Thermodynamics Review}

In a diffuse, single component ideal semirelativistic gas (Synge gas) of particles with rest mass $m$, the pressure is
\begin{equation}\label{eq.ideal}
    p = n k T
\end{equation}
where $n = \rho/m$, and the internal energy per unit volume is 
\begin{equation}\label{eq.syngeu}
    u = n k T  \left( \frac{K_3(\Theta^{-1})}{ K_2(\Theta^{-1})} 
    - 1 - \Theta \right) \frac{1}{\Theta}
\end{equation}
where $\Theta \equiv k T/(m c^2)$ \citep{chandra39, synge57}.  For $\Theta \ll 1$, $u = (3/2) n k T$, and for $\Theta \gg 1$, $u = 3 n k T$.  

For numerical applications the Bessel functions are expensive to evaluate. Easy-to-evaluate approximations have been developed by \cite{mathews71, service86, gammie98, mignone05, ryu06, sadowski17}.  The  approximation 
\begin{equation}
    u \approx n k T  \left( \frac{12 + 45 \Theta + 45 \Theta^2}{8 + 20 \Theta + 15 \Theta^2} \right)
\end{equation}
is good everywhere to better than $0.07\%$.

Recall that for a general equation of state there are three adiabatic exponents:
\begin{equation}\label{eq.allthegammas}
    \Gamma_1 \equiv \frac{d\ln p}{d\ln\rho}\biggr |_s, \qquad
    \frac{\Gamma_2}{\Gamma_2 - 1} \equiv \frac{d\ln p}{d\ln T}\biggr |_s, \qquad
    \Gamma_3 \equiv 
    \frac{d\ln T}{d\ln \rho}\biggr |_s.
\end{equation}
where the derivatives are evaluated at constant entropy $s$ \citep[e.g.][]{cox68}.  Notice that the three adiabatic exponents are ill defined for a two-temperature plasma.

Beginning with the first law, setting $dQ = 0$, and assuming $p = p(\rho,T)$ and $u = u(\rho,T)$, one can show that
\begin{equation}\label{eq.gengammaA}
    \Gamma_1 = 
        \frac{p_\rho \rho}{p} + 
        \frac{p_T (p + u - \rho u_\rho)}{p u_T}
\end{equation}
where the subscripts denote partial derivatives.

The ideal gas equation of state (\ref{eq.ideal}) implies the ratio of specific heats $\gamma = \Gamma_1 = \Gamma_2 = \Gamma_3$. For the Synge gas
\begin{equation}
    \gamma = 1 + \frac{\Theta^2 K_2^2}{(1 + 3\Theta^2) K_2^2 - 3 \Theta K_2 K_1 - K_1^2}.
\end{equation}
A serviceable approximation with the correct asymptotic behavior at small and large $\Theta \equiv k T/(m c^2)$, good everywhere to better than $0.4\%$, is
\begin{equation}\label{eq.syngefit}
    \gamma \approx \frac{5 + 20 \Theta + 24 \Theta^2}{3 + 15 \Theta + 18 \Theta^2}
\end{equation}
\citep[see also][]{sadowski17}.  
For a nonrelativistic ideal gas $\gamma = (N+2)/N$, where $N$ is the number of degrees of freedom, so a monatomic nonrelativistic ideal gas with three translational degrees of freedom has $N= 3$ and $\gamma = 5/3$.  For a relativistic ideal gas $\gamma = (N+1)/N$, because the equipartition theorem allots $k T$ to each degree of freedom, rather than $(1/2) k T$, when the energy is linear in the momentum, so a ideal relativistic gas with no internal degrees of freedom has $N = 3$ and $\gamma = 4/3$.

Notice that in the low-temperature limit $\Theta \ll 1$, $\gamma = (5/3)(1 - \Theta) + \mathcal{O}(\Theta^2)$. As the temperature increases the equation of state softens to $\gamma = (1/2) (5/3 + 4/3) = 3/2$ at $\Theta \approx 1/6$.  For electrons this is at $T_e \approx 10^9$K, and for protons at $T_p \approx 2 \times 10^{12}$K.  

\subsection{Astrophysical Setting}

Consider a disk composed of electrons and ions of mass $\mu m_p$.  The  scale height $H = h r$ at radius $r \equiv x G M/c^2$ from a black hole of mass $M$.  The temperature is $T = 6.5 \times 10^{12} h^2 (\mu/x)$K because vertical hydrostatic equilibrium implies $H = c_s/\Omega$, where $\Omega^2 \equiv G M/r^3$ and $c_s \equiv$ sound speed.  If $T_e = T_i/R$ then electrons are relativistic for $x < 6500 h^2 \mu/R$.  Then for $h \lesssim 1/2$, protons in a pure hydrogen gas never become relativistic outside the horizon.  Notice that temperatures are higher (by a factor of 4!) for a pure helium plasma \citep{wong22}, which may describe Sgr A* \citep{ressler18} where  helium-rich stellar winds \citep{martins07} probably feed the inner accretion flow.  Near the horizon, where the emission seen by EHT is produced, electrons are relativistic ($\Theta_e \sim 10^3 h^2 \mu/(R x)$) and ions are nonrelativistic ($\Theta_i \sim h^2/x$).  Then the electrons and ions have adiabatic indices $\gamma_e \approx 4/3$ and $\gamma_i \approx 5/3$.

\subsection{Combined Ion-Electron Gas}\label{subsec.iegas}

We consider four possible approaches to evaluating the adiabatic index. Not all are consistent with one another.  In all cases quasi-neutrality forces the densities of ions and electrons to evolve together (density changes occur on timescales long compared to the inverse plasma frequency).  

In the first case the electrons and ions equilibrate and have a single temperature.  In the second case the electrons and ions are decoupled and each component evolves independently and adiabatically.

In the first case the plasma is a mixture of ions (charge $Z = n_e/n_i$) and electrons, with adiabatic indices $\gamma_e$ and $\gamma_i$, and the two components are coupled by a relaxation process that forces {\em both to have the same temperature $T$}.  

Beginning with Equation (\ref{eq.sumu}), which assumes that $\gamma_i$ and $\gamma_e$ are constant, applying Equation (\ref{eq.gengammaA}), and setting $n_e = Z n_i$, find
\begin{equation}\label{eq.gammasingle}
\gamma = \frac{ \gamma_i (\gamma_e - 1) + \gamma_e (\gamma_i - 1) Z}
{\gamma_e - 1 + Z (\gamma_i - 1)}.
\end{equation}
This can be generalized to $\gamma_s = \gamma_s(T)$ ($s = e,i$) using Equation (\ref{eq.syngeu}), but nonconstant $\gamma_s$ is not the leading source of error here.

If $T$ is such that $\Theta_e > 1$ and $\Theta_i < 1$ (for a pure hydrogen gas, $10^9 {\rm K} < T < 2 \times 10^{12} {\rm K}$), then $\gamma_e \approx 4/3$, $\gamma_i \approx 5/3$, and
\begin{equation}\label{eq.adindeq}
\gamma \approx \frac{5 + 8Z}{3 + 6Z}.
\end{equation}
For pure hydrogen this yields the well-known result $\gamma = 13/9 = 1.44$ \citep{shapiro73}.  For pure helium $\gamma = 21/15 = 1.4$.  As shown below, however, this first case implicitly assumes that there is energy exchange between the ions and electrons.  Therefore $\gamma = 13/9$ is not appropriate to EHT sources.  

In the second case electrons and ions are decoupled and evolve independently with distinct temperatures $T_e$ and $T_i$.  If the plasma is compressed adiabatically then $p_s = \kappa_s \rho^{\gamma_s}$ and
\begin{equation}
p = p_i + p_e = \kappa_i \rho^{\gamma_i} + \kappa_e \rho^{\gamma_e}.  
\end{equation}
The adiabatic index can be found by evaluating $\Gamma_1$ in Equation (\ref{eq.allthegammas}) directly, holding $\kappa_e$ and $\kappa_i$ constant:
\begin{equation}
\gamma = \frac{p_i \gamma_i + p_e \gamma_e}{p_i + p_e},
\end{equation}
or, using the ideal gas law,
\begin{equation}\label{eq.gammaad}
\gamma = \frac{\gamma_i T_i  + \gamma_e Z T_e}{T_i + Z T_e} =
\frac{\gamma_i R + \gamma_e Z}{R + Z}.
\end{equation}
For $R \gg Z$, $\gamma \approx \gamma_i + (Z/R) (\gamma_e - \gamma_i) + \mathcal{O}(Z/R)^2$.  If $R \gg Z$, as is assumed in most GRMHD models of EHT sources, then for a single-fluid model the best approximation is $\gamma \approx 5/3$.  

In the third case suppose one wants to assign an effective equation of state in which $p \approx (\gamma-1) u$, as in \cite{ressler15, sadowski17}.  Then
\begin{equation}\label{eq.effectiveEOS}
    \gamma = 1 + \frac{p}{u} = 1 + \frac{(\gamma_i - 1)(\gamma_e - 1) (R + Z)}{(\gamma_i - 1) Z + (\gamma_e - 1) R}.
\end{equation}
For $R \gg Z$, $\gamma \approx \gamma_i + (Z/R)(\gamma_e - \gamma_i) (\gamma_i - 1)/(\gamma_e - 1) + \mathcal{O}(Z/R)^2$. In this case the best approximation for a single-fluid model for EHT sources is still $\gamma \approx 5/3$.  

In the fourth case suppose that one requires that as the fluid evolves there is a process that fixes $T_i = R T_e$.  Assume a single effective temperature $T \propto T_i$ (it does not matter what the constant of proportionality is) then one recovers Equation (\ref{eq.effectiveEOS}) exactly and again the best approximation for a single fluid model is $\gamma \approx 5/3$.  

\section{Two Component Model}\label{sec.two}

Since there is likely some coupling between ions and electrons, albeit weak, one might wonder if the discussion above is complete.  Here we introduce a model for evolution of $T_e$ and $T_i$ that includes finite coupling.  

Begin with the first law of thermodynamics in the form
\begin{equation}\label{eq.firstlaw}
\frac{Du}{Dt} = \rho T \frac{Ds}{Dt} + \frac{u + p}{\rho} \frac{D\rho}{D t}
\end{equation}
where $D/Dt$ is the convective derivative (a hydrodynamic translation of the usual thermodynamic  ``$d$'').   Rewrite this as 
\begin{equation}\label{eq.firstlawp}
\frac{Du}{Dt} = Q + \frac{\gamma n k T}{\gamma - 1} \frac{D\ln \rho}{D t} = Q +  \frac{\gamma n k T}{\gamma - 1}\frac{1}{\tau_{comp}} 
\end{equation}
which defines $Q$, the heating/cooling rate per unit volume, and $\tau_{comp}$, a compression timescale.  Notice that $\tau_{comp} < 0$ in an expanding plasma.

There are three contributions to $Q$: ion-electron energy exchange, turbulent dissipation, and radiative cooling.  

The ion-electron energy exchange heating/cooling rate per unit volume is
\begin{equation}
Q^{ie}_e = -\mathcal{C}_e (T_e - T_i)/\tau_{ie}
\end{equation}
\begin{equation}
Q^{ie}_i = -\mathcal{C}_i (T_i - T_e)/\tau_{ie}
\end{equation}
where $\mathcal{C}_s$ is dimensional and $\tau_{ie}$ is an energy exchange timescale.  Energy conservation requires  $Q^{ie}_e = - Q^{ie}_i$ so $\mathcal{C}_i = \mathcal{C}_e = \mathcal{C} > 0$.  Our model, which could include Coulomb interactions but is also meant to include ion-electron energy exchange mediated by electromagnetic field fluctuations, is $\mathcal{C} = (u_e + u_i)/T_{ie}$, where $T_{ie}$ is a suitable symmetric average of $T_e, T_i$.   In the regime of interest, where electrons are relativistic and ions nonrelativistic, $T_{ie} \approx T_e$ for Coulomb scattering \citep{stepney83}, so we assume $T_{ie} = T_e$ from now on.   

The turbulent dissipation heating rate per unit volume is 
\begin{equation}
Q^{+}_e = f_e Q_{tot} 
\end{equation}
\begin{equation}
Q^{+}_i = f_i Q_{tot}
\end{equation}
where $f_e + f_i = 1$, $Q_{tot} = (u_i + u_e)/\tau_{diss}$, and $\tau_{diss}$ is a dissipation timescale.   

For completeness the radiative cooling rate per unit volume is 
\begin{equation}
    Q^{-}_e = -\frac{u_e}{\tau_{cool}}
\end{equation}
\begin{equation}
    Q^{-}_i = 0
\end{equation}
where $\tau_{cool} > 0$ is a characteristic radiative cooling timescale. 

The total heating rate per unit volume is $Q_s = Q^{ie}_s + Q^+_s + Q^-_s$.

Expanding the first law, 
\begin{equation}\label{eq.dlnTspec}
\frac{D\ln T_s}{Dt} = \frac{\gamma_s - 1}{k n_s T_s} Q_s + \frac{\gamma_s - 1}{\tau_{comp}}
\end{equation}
where $D\ln X/Dt = (1/X) DX/Dt$. 
This completes specification of the model.

\section{Strong and Weak Coupling Limits}\label{sec.limits}

Here we show that the two-component model produces Equation (\ref{eq.gammaad}) in the limit that ion-electron coupling is weak ($\tau_{ie} \rightarrow \infty$) and Equation (\ref{eq.gammasingle}) in the limit that ion-electron coupling is strong ($\tau_{ie} \rightarrow 0$).  To do this we evaluate $\gamma$ using 
\begin{equation}\label{eq.gengamma}
    \gamma = \frac{D\ln p}{Dt}\, / \, \frac{D\ln \rho}{Dt}.
\end{equation}

First the weak-coupling limit.  If $\tau_{ie} \rightarrow \infty$ then $Q^{ie}_s \rightarrow 0$.  Using $p d\ln p = p_i d\ln p_i + p_e d\ln p_e$, $d\ln p_s = d\ln \rho + d\ln T_s$, and setting $Q = 0$
\begin{equation}
\gamma = \frac{\gamma_i T_i  + \gamma_e Z T_e}{T_i + Z T_e} =
\frac{\gamma_i R + \gamma_e Z}{R + Z},
\end{equation}
which is identical to Equation (\ref{eq.gammaad}).   

Next, the strong-coupling limit, where $\tau_{ie}$ is small.  Since $Q^{ie} \propto (T_e - T_i)/\tau_{ie}$, strong coupling will produce nearly equal ion and electron temperatures.  There must still be a small temperature difference, however, to drive energy exchange between species; for equal temperatures $Q^{ie} = 0$ and a temperature difference would be created by compression.  Taking $\tau_{ie} \sim \epsilon \tau_{comp}$ and assuming that $\epsilon \ll 1$, that $T_i - T_e \sim \epsilon T_i$ and that $T_i - T_e$ is approximately constant, then Equation (\ref{eq.dlnTspec}) at  lowest order in $\epsilon$ implies
\begin{equation}
    T_i - T_e \approx T_i \, \frac{Z (\gamma_e - 1)(\gamma_i - \gamma_e) (\gamma_i - 1)}{(\gamma_e - 1 + Z (\gamma_i - 1) Z)^2 }\left(\frac{\tau_{ie}}{\tau_{comp}}\right) + \mathcal{O}(\epsilon^2).
\end{equation}
Using this to evaluate Equation (\ref{eq.gengamma}) to lowest order in $\epsilon$, 
\begin{equation}
    \gamma = \frac{ \gamma_i (\gamma_e - 1) + \gamma_e (\gamma_i - 1) Z}
{\gamma_e - 1 + Z (\gamma_i - 1)},
\end{equation}
which is identical to Equation (\ref{eq.gammasingle}).  

The appropriate adiabatic index is $5/3$, then, if ion-electron coupling is weak, $\Theta_i \ll 1$, and $R \gg Z$.  More intuitively, if the electrons are cool enough that they do not contribute to the pressure then they can have no effect on the change in pressure under compression, whether they are relativistic or not.

Finally, notice that we can evaluate $\gamma$ in the strong coupling limit for the fourth case described in subsection \ref{subsec.iegas}, where an  unspecified relaxation process maintains $T_i \approx R T_e$.  In this case we recover Equation (\ref{eq.effectiveEOS}). 

\section{Equilibrium Temperature Ratio}\label{sec.eqrat}

What is the temperature ratio $R$?  On the observational side, models of EHT sources ``paint'' the electrons using an $R$ prescription that depends on local conditions, most commonly the $R(\beta)$ prescription of \cite{moscibrodzca16}:\begin{equation}
    R = R_{low}\frac{1}{1 + \beta^2} + R_{high} \frac{\beta^2}{1 + \beta^2}.
\end{equation}
Here $\beta \equiv p/(B^2/(8\pi)$ is the ratio of gas to magnetic pressure.  These models do not fare well in comparison to data when $R_{low} = R_{high} = 1$ \citep{EHTC_2019_5, EHTC_2021_8, EHTC_2022_5}.  In this sense models with $R \gg 1$ are observationally well motivated.  On the theoretical side, workers have estimated the ratio $Q^+_i/Q^+_e$ \citep[e.g.][]{quataert99, sharma07, howes10, kawazura19}.  These estimates find $Q^+_i/Q^+_e > 1$ when $\beta \gtrsim 1$.  Since $Q^+_i/Q^+_e > 1$ increases $R$, models with $R > 1$ are also theoretically well motivated. 

It is not immediately obvious how $Q^+_i/Q^+_e$ (a ratio of heating rates) and $R$ (a ratio of temperatures) are related.  Using the two-component model we will show that there is a direct connection between the two when the temperature ratio is in equilibrium, i.e. $R = const.$  We ignore electron cooling, which always increases $R$.

Evolution of $R$ follows
\begin{equation}
\frac{D\ln R}{Dt} = \frac{D\ln T_i}{Dt} - \frac{D\ln T_e}{Dt} = 
	\frac{\gamma_i - 1}{k n_i T_i} Q_i - \frac{\gamma_e - 1}{k n_e T_e} Q_e + \frac{\gamma_i - \gamma_e}{\tau_{comp}}.
\end{equation}
Expanding this and setting $\gamma_i = 5/3, \gamma_e = 4/3, T_{ie} = T_e,$ and $ Z = 1$,
\begin{equation}\label{eq.Revolve}
\frac{D\ln R}{Dt} = 
	\frac{1}{\tau_{comp}}(\gamma_i - \gamma_e)+ 
	\frac{1}{\tau_{diss}} \left( f_i (1 + \frac{2}{R}) - f_e (1 + \frac{R}{2}) \right) -
	\frac{1}{\tau_{ie}} \frac{(R - 1) (R + 2)^2}{2 R} 
\end{equation}
Suppose that a parcel of plasma reaches equilibrium  ($D\ln R/D t = 0$) and take the weak coupling limit.  Then to lowest order in $\epsilon \sim 1/R \sim f_e$,
\begin{equation}
R \simeq \frac{1}{f_e} \left(2 + \frac{2}{3} \frac{\tau_{diss}}{\tau_{comp}}\right) + \mathcal{O}(1).
\end{equation}
Since in models of turbulent dissipation \citep[e.g.][]{quataert99, sharma07, howes10, kawazura19} $f_e$ is usually found to decrease as $\beta$ increases, this motivates $R(\beta)$ models like that of \cite{moscibrodzca16}.

Outside the weak coupling limit $\tau_{ie}$ is comparable to other timescales and $R$ will be smaller.  The general equilibrium solution of for $R$ is complicated but easy to obtain from Equation (\ref{eq.Revolve}), which is a quartic in $R$.  Electron cooling, neglected here, will only increase $R$.  

\section{Conclusion}\label{sec.conc}

Numerical fluid models of low luminosity black hole accretion flows must specify an equation of state.  Best practice would be to introduce separate internal energy equations for ions and electrons, explicitly model heating, electron cooling, and ion-electron energy transfer \citep[e.g.][]{ressler15, ryan18}, and introduce electron pressure in the energy-momentum equations \citep{sadowski17, liska24}.  This comes at the cost of additional parameters associated with the heating, cooling, and transfer models.  For reasons of simplicity and computational expense this is not usually done.  Instead modelers evolve a total internal energy $u = u_e + u_i$ and adopt an effective  equation of state $p = (\gamma - 1) u$.  The literature includes models with $\gamma = 5/3, 13/9$, and $4/3$.  Out of these options we have argued that the best choice is $\gamma = 5/3$.

More accurately, the adiabatic index is slightly less than $5/3$.  There is a  reduction in adiabatic index related to finite $\Theta_i$, $\gamma \approx (5/3) (1 - \Theta_i)$ (Equation [\ref{eq.syngefit}]), and a slight reduction from the electrons, $\gamma \approx 5/3 - Z/(3 R)$ (Equation [\ref{eq.gammaad}]).  These small reductions are consistent with the reductions in adiabatic index reported in \cite{liska24}.  

It is well known that in spherical accretion $\gamma = 5/3$ is a singular case, with a sonic point appearing only due to relativistic effects close to the horizon \citep[][Appendix G]{shapiro83}.  This happens because when $\gamma = 5/3$ the gravitational binding energy, internal energy, and kinetic energy of the plasma all scale in the same way with radius in the Newtonian regime and the spherical accretion solution is then self-similar.  The sensitivity of the flow to $\gamma$ near $\gamma = 5/3$ is also evident in the \cite{narayan94} ADAF model.  It is not yet known whether there are differences between GRMHD simulations with $\gamma = 5/3$ and $\gamma = 5/3 - \epsilon$.

Simulations that evolve the ion and electron internal energy separately will differ from those that using an effective equation of state with $\gamma = 5/3$.  To estimate where the inconsistency is greatest we temporarily adopt an $R(\beta)$ models with $R_{\rm low} = 1$. Then where $\beta \lesssim 1$, $R \sim 1$ and the true adiabatic index according to Equation (\ref{eq.gammaad}) is $3/2$, as long as the temperature is between $10^9$K and $10^{12}$K.  The error in the GRMHD evolution introduced by an effective equation of state approximation is mitigated in the $R(\beta)$ model by the dominance of magnetic pressure over gas pressure ($\beta \lesssim 1$) in the regions where the error is largest.   

Our two-component model has implications for the electron temperature in jets and outflows.  In the model the difference in ion and relativistic electron adiabatic indices alone drives changes in $T_i/T_e$ where the plasma is expanding or contracting.  In jets and outflows expansion drives a reduction in $R$ because in the absence of heating or cooling, $R \propto \rho^{1/3}$.  Expansion also drives a reduction in $T_e$, but this decrease will be smaller than predicted by an $R(\beta)$ model.

Our two-component model also has implications for the connection between synchrotron emissivity fluctuations and density fluctuations. The thermal synchrotron emissivity $j_\nu = j_\nu(\rho, B, T_e)$.  For an adiabatic density fluctuation $\delta \rho$ the change in emissivity due to temperature fluctuations alone is $\delta j_\nu \approx \delta T_e \partial j_\nu/\partial T_e$.  In the weak coupling limit $\delta T_e = (\gamma_e - 1) T_e (\delta \rho/\rho)$.  If $R$ is fixed, as in the $R(\beta)$ prescription away from $\beta = 1$, $T_e = T_i/R$, $\delta T_i = (\gamma_i - 1) T_i (\delta \rho/\rho)$, and $\delta T_e = (\gamma_i - 1) T_e (\delta \rho/\rho)$.  Thus in the weak coupling limit the electron temperature fluctuation is smaller than in a fixed $R$ model by a factor of $(\gamma_e - 1)/(\gamma_i - 1) \approx 1/2$.  This may partially explain the excess variability seen in GRMHD models of Sgr A* \citep{EHTC_2022_5}.

\acknowledgements

This work was supported by NSF grants AST 17-16327 (horizon), OISE 17-43747, and AST 20-34306.  This work was supported in part by the IBM Einstein Fellow Fund at
the Institute for Advanced Study.  I thank Le\'on Sosapanta Salas for stimulating discussions, and George Wong, Stu Shapiro, Nick Conroy, Abhishek Joshi, Michi
Baub\"ock, Ramesh Narayan, Andrew Chael, and an anonymous referee for comments that improved the paper.

\newpage
\bibliographystyle{aasjournal}
\bibliography{references}

\end{document}